\documentclass[preprint2]{aastex6}

\usepackage{bm}
\usepackage{color}
\usepackage{amsmath}
\usepackage{natbib}
\usepackage{multirow}
\usepackage{here}
\begin{document}


\title{CONVECTIVE VELOCITY SUPPRESSION VIA THE ENHANCEMENT OF SUBADIABATIC LAYER: \\
  ROLE OF THE EFFECTIVE PRANDTL NUMBER}


\author{Y.Bekki\altaffilmark{1}, H.Hotta\altaffilmark{2} and T.Yokoyama\altaffilmark{1}}
\affil{$^{1}$Department of Earth and Planetary Science, The University of Tokyo, 7-3-1 Hongo, Bunkyo-ku, Tokyo 113-0033, Japan}
\affil{$^{2}$Department of Physics, Chiba University, 1-33 Yayoi-cho, Inage-ku, Chiba 263-8522, Japan}







\altaffiltext{1}{bekki@eps.s.u-tokyo.ac.jp}

\begin{abstract}
  
  It has recently been recognized that the convective velocities achieved in the current solar convection simulations might be over-estimated.
  The newly-revealed effects of the prevailing small-scale magnetic field within the convection zone may offer possible solutions to this problem.
  The small-scale magnetic fields can reduce the convective amplitude of small-scale motions through the Lorentz-force feedback, which concurrently inhibits the turbulent mixing of entropy between upflows and downflows.
  As a result, the effective Prandtl number may exceed unity inside the solar convection zone.
  In this paper, we propose and numerically confirm a possible suppression mechanism of convective velocity in the effectively high-Prandtl number regime.
  If the effective horizontal thermal diffusivity decreases (the Prandtl number accordingly increases), the subadiabatic layer which is formed near the base of the convection zone by continuous depositions of low entropy transported by adiabatically downflowing plumes is enhanced and extended.
  The global convective amplitude in the high-Prandtl thermal convection is thus reduced especially in the lower part of the convection zone via the change in the mean entropy profile which becomes more subadiabatic near the base and less superadiabatic in the bulk.

\end{abstract}




\section{Introduction} \label{sec:intro}

Thermal convection plays a critical role inside the solar interior; it transports heat while powering dynamos.
For investigating the convective properties and the dynamo processes occurring inside the convection zone, three dimensional (3D) full-spherical hydrodynamic (HD) or magnetohydrodynamic (MHD) simulations have been conducted \citep[e.g.,][]{glatzmaier1984,miesch2000,brun2004}.
However, it has been recently recognized that the recent 3D solar convection simulations which aim to achieve more realistic regimes with higher resolutions tend to over-estimate the convective amplitudes, which is commonly known as the convective conundrum \citep[see][for details]{hanasoge2016review}.

Recent local helioseismic observations by \citet{hanasoge2012} inferred the large-scale convective amplitude which is about two orders of magnitude smaller than the value typically predicted by mixing-length model or obtained in global HD simulations \citep{miesch2008}.
It should be noted that the helioseismic estimates depend on the inversion technique employed, for \citet{greer2015} instead found much larger flow velocities that are consistent with the theoretical models.
Indeed, realistic surface convection simulations also reported the large-scale velocity power an order of magnitude larger than photospheric observations, also suggesting that the numerical simulations are over-powering the convection larger than supergranular scales \citep{lord2014}.

Another evidence comes from the problem related with the differential rotation profile.
Recently, many high-resolution global simulations with solar parameters, such as the solar luminosity $L_{\odot}$ and the rotation rate $\Omega_{\odot}$, reported that the anti-solar differential rotations with faster rotating pole and slowly rotating equator were unexpectedly achieved \citep{gastine2013,kapyla2014,fan2014,hotta2015b,karak2015}.
It is generally believed that this anti-solar differential rotation profile is attributed to a large Rossby number $\mathrm{Ro} \equiv v/(2\Omega l)$, which measures the effect of convection with respect to the rotation, where $\Omega$, $v$ and $l$ are the rotation rate, typical convective velocity and length scale, respectively.
If the simulated convective velocities are over-estimated, the Rossby number becomes large and then the angular momentum is transported radially inward by the turbulent Reynolds stress, which finally results in producing the anti-solar differential rotation profile \citep{featherstone2015}.
We believe therefore that if the problem of too fast convection obtained in the recent high-resolution simulations is solved, it will highly contribute to alleviating the problem of anti-solar differential rotation simultaneously. 

The problem described so far suggests the existence of some physical processes that the recent global (M)HD simulations neglected or just could not capture well enough.
A possible solution to the problem is small-scale magnetic fields generated by small-scale dynamos \citep[e.g.,][]{brandenburg2005}.
Small-scale dynamos can operate when the magnetic Reynolds number $\mathrm{Rm} \equiv vl/\eta$ exceeds the critical value and can generate magnetic fields on scales smaller than the energy-carrying scale of the turbulent flow.
The effects of small-scale magnetic fields are not included in the HD simulations, and even for MHD case, it is highly likely that most global simulations have insufficient resolutions to achieve large enough $\mathrm{Rm}$ to resolve the inertial range of the turbulence, which is essential for the small-scale dynamo.
The importance of the small-scale dynamo near the surface has been widely recognized in explaining the observed mixed-polarity field in the photosphere \citep{cattaneo1999,vogler2007,rempel2014} where the flows are essentially non-helical due to a sufficiently large $\mathrm{Ro}$.
On the other hand, in the deeper convection zone where large-scale dynamos can be easily excited by helical turbulence, little attention has been paid so far to the effects of the small-scale dynamo.

Recently, however, \citet{hotta2015,hotta2016} conducted high-resolution MHD simulations of the solar convection zone with and without rotation and finally demonstrated that the small-scale dynamos can also be efficiently excited throughout the convection zone even when the rotational effects are included.
It is shown that the magnetic energy becomes close to equipartition or even superequipartition with the convective kinetic energy at small scales.
The Maxwell stress related to these superequipartition magnetic fields is found to act similarly to a viscous stress for convective motions.
As a result, the convective velocities are suppressed via the Lorentz force feedback by $50\%$ in the highest-resolution MHD simulation of \citet{hotta2015}.
Although, the small-scale dynamo is still not saturated in the sense that it does not converge in resolution, the suppression of convective velocities directly through the Lorentz-force seems not enough for explaining the huge discrepancy between observations and simulations. 

Another effect of the small-scale magnetic field which prevails within the convection zone was pointed out and first investigated by \citet{omara2016}.
The small-scale magnetic field not only inhibits shearing motions but also reduces the horizontal turbulent mixing of the entropy perturbations between warm upflows and cold downflows \citep{hotta2015}.
In other words, the effective viscosity $\nu_{\mathrm{SGS}}$ is enhanced while the effective thermal diffusivity $\kappa_{\mathrm{SGS}}$ is reduced, leading to an increase in the effective Prandtl number $\mathrm{Pr_{eff}} \equiv \nu_{\mathrm{SGS}} /\kappa_{\mathrm{SGS}}$ which can be larger than unity.
\citet{omara2016} conducted a set of solar convection simulations in which $\kappa_{\mathrm{SGS}}$ is decreased while $\nu_{\mathrm{SGS}}$ is kept fixed (and thus $\mathrm{Pr_{eff}}$ is increased).
It was shown that the convective velocity systematically decreases as $\mathrm{Pr_{eff}}$ is increased.
They attribute this decrease in convective velocities to the large thermal contents of downflowing plumes originating from a strongly-superadiabatic surface layer, which is formed by their vertical thermal conductive-type upper boundary condition.
In short, convection becomes weak in high-$\mathrm{Pr_{eff}}$ regime so that the fixed solar luminosity $L_{\odot}$ can be transported outward with larger thermal contents possessed by warmer upflows and colder downflows.

Let us distinguish these two effects of the small-scale magnetic field by referring to the former one (suppression by the Lorentz force) as a dynamical effect and the latter (suppression via the change of $\mathrm{Pr_{eff}}$) as a thermal effect.
Through the dynamical effect of the Lorentz force, the kinetic energy of the convective flows can be reduced via the energy exchanges between the dynamo-generated magnetic energy.
The importance of the thermal effect via the decrease in $\kappa_{\mathrm{SGS}}$ is significant because it describes that the reduction in the kinetic energy can also result from an increase in the internal energy.
Since the internal energy is far larger than both the kinetic and the magnetic energy in the deep convection zone owing to a very small Mach number ($\mathrm{M}^{2} \sim 10^{-7}$) and a very large plasma beta ($\beta\sim 10^{6}$) \citep{ossendrijver2003}, the thermal effect will offer another possibility to resolve the huge discrepancies between simulations and observations.

\begin{figure*}[]
\figurenum{1}
\plotone{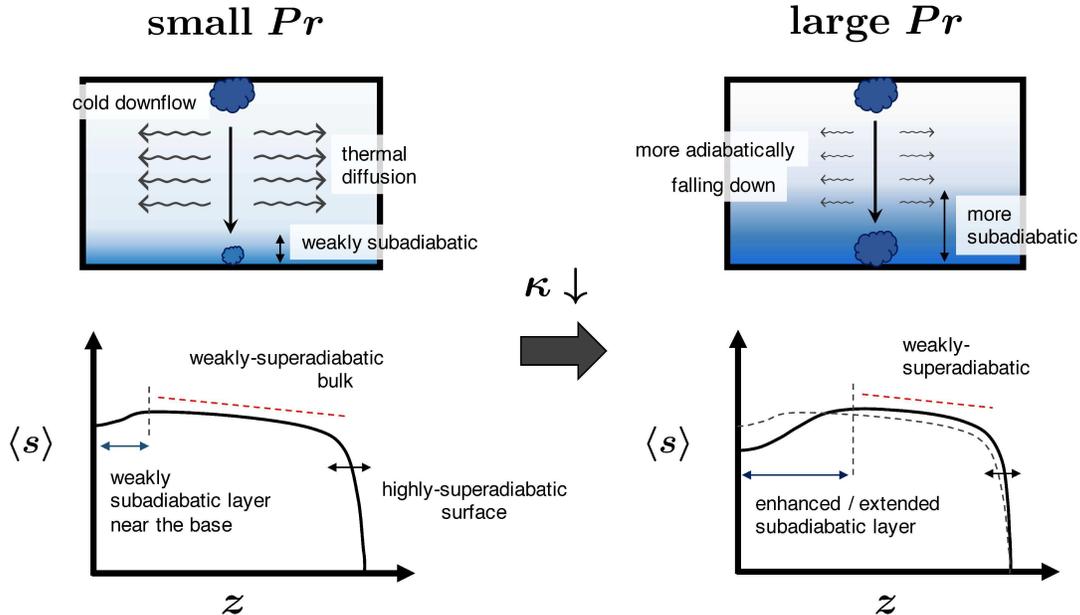}
\caption{Schematic illustration of the velocity suppression mechanism in low-$\kappa$ (high-$\mathrm{Pr}$) thermal convection. The upper panels show the evolutions of cold downflowing plumes with the expected background subadiabatic stratification over-plotted. The size of a cloud represents the amplitude of entropy perturbation of downflows. For simplicity of the figures, highly superadiabatic surface layers are neglected. Lower panels show the expected mean entropy profiles for both small and large $\mathrm{Pr}$ cases.} \label{fig:mech1}
\end{figure*}

In this paper, we propose a velocity suppression mechanism that can be achieved in high-$\mathrm{Pr_{eff}}$ regime which is distinct from what was described in the previous study of \citet{omara2016}.
We further confirm the suppression mechanism by conducting a set of HD simulations of Cartesian box where the effects of small-scale magnetic fields are modeled as sub-grid-scale (SGS) diffusivities, $\nu_{\mathrm{SGS}}$ and $\kappa_{\mathrm{SGS}}$.
Previous small-scale dynamo simulations revealed that the horizontal root-mean-square (rms) velocity is more decreased than the vertical one, which leads to a main reduction of a ``horizontal'' heat transport by the turbulence \citep{hotta2015}.
We therefore introduce anisotropy in the SGS thermal diffusion and examine a dependence of the convective amplitudes on each of the horizontal and vertical SGS thermal diffusivities.

The organization of the paper is as follows.
In Section \ref{sec:two}, we describe a possible convective velocity suppression mechanism.
In Section \ref{sec:three}, our numerical model is explained. The simulation results for both isotropic and anisotropic thermal diffusion cases are presented in Section \ref{sec:four}.
We discuss implications on the solar convection zone dynamics in Section \ref{sec:five}, and the conclusions are summarized at last in Section \ref{sec:six}.

\section{Effects of the Decrease in the Effective Thermal Diffusivity} \label{sec:two}

In this section, we examine possible effects of a decrease in the effective thermal diffusivity which can be brought about by small-scale magnetism.
We argue here that the convective velocity can be suppressed via the enhancement of a weakly subadiabatic layer near the base of the convection zone.

Figure \ref{fig:mech1} shows a schematic illustration of this process which is explained as follows.
If thermal diffusivity $\kappa$ is reduced, cold downflows become able to fall down without losing their large thermal contents via the thermal diffusion so that the low entropy fluids are transported more adiabatically to the bottom.
Note that continuous depositions of the low entropy near the bottom may naturally form a weakly-subadiabatic (convectively-stable) layer.
The formation of this subadiabatic layer near the base would be therefore enhanced and extended in low-$\kappa$ (and thus high-$\mathrm{Pr}$) regime.
The influence of the mean entropy stratification on convection is evaluated by superadiabaticity $\delta\equiv\nabla-\nabla_{\mathrm{ad}}$ with $\nabla\equiv d\ln T / d \ln p$.
The final value of subadiabaticity ($\delta <0$) near the base achieved in a statistically-stationary convection would be determined by a balance between a continuous supply of low entropy by downflows and suppressions of downflows which in turn limits the amount of low entropy supply.
Note that if $\kappa$ is relatively large, a relaxation effect by the vertical thermal conduction also enters this balance.
The enhanced and extended subadiabatic layer in the lower convection zone may have a considerable impact on suppressing global convective amplitudes: it not only suppresses the vertical motions locally through the buoyant deceleration but it can also limit the global convective mode to shallower scales \citep{kapyla2017b} which then leads to a reduction of the global convective velocity.

In order to distinguish several thermal effects operating in the low-$\kappa$ regime, it is instructive to define the horizontal and vertical thermal diffusivities, $\kappa^{\mathrm{H}}$ and $\kappa^{\mathrm{V}}$ and discuss the thermal diffusion in each of the directions.
We consider that the decrease in $\kappa^{\mathrm{H}}$ is essential for downflows to retain their large amount of entropy deficits and to achieve a mean entropy stratification which is more subadiabatic near the base.
However, $\kappa^{\mathrm{V}}$ also has several important effects that can influence the proposed velocity suppression mechanism.
Near the top and bottom boundaries, a decrease in $\kappa^{\mathrm{V}}$ prohibits the vertical relaxation of the mean entropy stratification, leading to a more subaidiabatic base and more superadiabatic surface.
Moreover, near the surface layer where convection is continuously driven, the decrease in $\kappa^{\mathrm{V}}$ has another significant effect in increasing the Rayleigh number ($\mathrm{Ra}\propto 1/\kappa^{\mathrm{V}}$) and thus increasing the degree of turbulence there.

In the previous numerical studies of compressible thermal convection with $\mathrm{Pr}$ larger than unity, the thermal diffusion has been treated isotropically \citep{warnecke2014,omara2016,kapyla2017a}, and thus, all of the thermal effects described above are simultaneously included.
However, recent MHD simulations revealed that the small-scale magnetism tends to make the small-scale motions highly anisotropic, suggesting that the SGS turbulent transport coefficients should also be anisotropic:
More precisely, a main decrease in $\kappa^{\mathrm{H}}$ is inferred \citep{hotta2015}.
In order to distinguish the different thermal effects and to investigate the influence of $\kappa^{\mathrm{H}}$ on the proposed velocity suppression mechanism, the anisotropy in the thermal diffusion is necessary.
We thus conduct a set of high-$\mathrm{Pr}$ convection simulations to numerically confirm that the enhanced subadiabatic layer can suppress the global convective amplitude with anisotropic thermal diffusion included.

\section{Numerical Model} \label{sec:three}

\subsection{Basic Equations} \label{subsec:3-1}

In this investigation, we solve the three-dimensional hydrodynamic equations in a simplified system with Cartesian coordinate $(x, y, z)$.
In our definition, $z$-axis is directed vertically upward, anti-parallel to the gravity.
The basic equations consist of the equation of continuity, the equation of motion, the equation of entropy, and the equation of state,
\begin{eqnarray}
  \frac{\partial \rho_{1}}{\partial t}&=&-\nabla\cdot(\rho_{0}\bm{v}),\label{eq:conti} \\
  \frac{\partial \bm{v}}{\partial t}&=&-\bm{v}\cdot\nabla\bm{v}-\frac{\nabla p_{1}}{\rho_{0}}-\frac{\rho_{1}}{\rho_{0}}g\bm{e}_{z}+\frac{1}{\rho_{0}}\nabla\cdot\bm{\Pi}, \label{eq:motion} \\
  \frac{\partial s}{\partial t}&=&-\bm{v}\cdot \nabla s + \frac{1}{\rho_{0}T_{0}}\nabla\cdot(\rho_{0}T_{0} \bm{\kappa}\cdot\nabla s) \nonumber \\
  && \ +\frac{\gamma -1}{p_{0}}(Q_{\mathrm{vis}}+Q_{\mathrm{heat}}+Q_{\mathrm{cool}}), \label{eq:ent} \\
  p_{1}&=&p_{0}\left(\gamma \frac{\rho_{1}}{\rho_{0}}+ s \right).
\end{eqnarray}
Here, $\rho_{0}(z)$, $p_{0}(z)$, and $T_{0}(z)$ denote time-independent reference state values of density, pressure, and temperature, respectively.
The reference state is assumed to be in an adiabatically-stratified hydrostatic equilibrium.
The thermodynamic variables with subscript 1, $\rho_{1}$ , $p_{1}$ , and $T_{1}$ represent perturbations from the reference state that are small compared with background values so that the equation of continuity and the equation of state are linearized.
Note that the entropy $s$ is normalized by the specific heat capacity at constant volume $c_{\mathrm{v}}$ and the ideal gas is assumed for the ratio of specific heats, $\gamma = 5/3$.
$g \ (>0)$ denotes the gravitational acceleration and is assumed to be constant in space.

The reference state quantities are given in the following forms that are identical to \citet{fan2003},
\begin{eqnarray}
  \rho_{0}(z)&=&\rho_{\mathrm{r}}\left[1-\frac{z}{(1+m)H_{\mathrm{r}}} \right]^{m}, \\
  p_{0}(z)&=&p_{\mathrm{r}}\left[1-\frac{z}{(1+m)H_{\mathrm{r}}} \right]^{1+m}, \\
  T_{0}(z)&=&T_{\mathrm{r}}\left[1-\frac{z}{(1+m)H_{\mathrm{r}}} \right], \\
  H_{0}(z)&=&\frac{p_{0}}{\rho_{0} g},
\end{eqnarray}
where $\rho_{\mathrm{r}}$, $p_{\mathrm{r}}$, $T_{\mathrm{r}}$, and $H_{\mathrm{r}}$ denote the values of $\rho_{0}$, $p_{0}$, $T_{0}$, and the pressure scale height $H_{0}$ evaluated at the bottom $z = 0$, respectively.
Polytropic index $m$ takes an adiabatic value $m=1/(\gamma-1)$.

\subsection{Sub-Grid-Scale Diffusivities} \label{subsec:3-2}

$\bm{\Pi}$ denotes the viscous stress tensor and $Q_{\mathrm{vis}}$ is the amount of dissipated energy which is converted from kinetic energy into internal energy.
They are given by,
\begin{eqnarray}
  \Pi_{ij}&=&\rho_{0}\nu\left[\frac{\partial v_{i}}{\partial x_{j}}+\frac{\partial v_{j}}{\partial x_{i}}-\frac{2}{3}(\nabla\cdot\bm{v})\delta_{ij} \right], \\
  Q_{\mathrm{vis}}&=&\sum_{i,j} \Pi_{ij}\frac{\partial v_{i}}{\partial x_{j}},
\end{eqnarray}
where $\nu$ and $\delta_{ij}$ are the coefficient of kinetic viscosity and the Kronecker delta, respectively.
Note that, for the sake of simplicity, we try to model the Maxwell stress on scales smaller than the grid resolution (sub-grid-scale) $M_{ij}\equiv\overline{b_{i}'b_{j}'}/4\pi$ by the viscous stress on the resolved scale $\Pi_{ij}$.
Here, $\bar{\ }$ represents an ensemble average on sub-grid-scale so that $\overline{\bm{v}}=\bm{v}$, $\overline{\bm{B}}=0$.
Therefore, the coefficient $\nu$ should be regarded as a diffusive coefficient of an effective viscosity which mimics the small-scale magnetic tension force.

On the other hand, the thermal diffusivity $\kappa$ is regarded as a sub-grid-scale eddy diffusivity reflecting the heat transport by unresolved turbulent motions that are subject to a strong Lorentz-force feedback of superequipartition small-scale magnetic fields.
Small-scale dynamo simulations of \citet{hotta2015} showed that the small-scale magnetism has a significant effect in suppressing the ``horizontal'' rms velocity, and as a result, the horizontal turbulent heat transport is greatly reduced.
This means that the effective thermal diffusivity $\kappa$ should be suppressed in a horizontal direction when we take into account the unresolved turbulent motions that are magnetized.
We, therefore, introduce the anisotropic sub-grid-scale thermal diffusion by expressing the thermal diffusivity tensor $\bm{\kappa}$ as,
\begin{eqnarray}
  \bm{\kappa}=\left[
    \begin{array}{rrr}
      \kappa^{\mathrm{H}} & & 0 \\
      & \kappa^{\mathrm{H}} & \\
      0 & & \kappa^{\mathrm{V}}
      \end{array}
    \right].
\end{eqnarray}
Here, $\kappa^{\mathrm{H}}$ and $\kappa^{\mathrm{V}}$ are the horizontal and vertical thermal diffusivities, respectively.
Both $\nu$ and $\kappa^{\mathrm{H,V}}$ are assumed to have the same height dependence,
\begin{eqnarray}
  \frac{\nu(z)}{\nu_{\mathrm{r}}}=\frac{\kappa^{\mathrm{H}}(z)}{\kappa_{\mathrm{r}}^{\mathrm{H}}}
  =\frac{\kappa^{\mathrm{V}}(z)}{\kappa_{\mathrm{r}}^{\mathrm{V}}}=\left(\frac{\rho_{0}(z)}{\rho_{\mathrm{r}}}\right)^{-1/2}. \label{eq:v-depend}
\end{eqnarray}
The same functional form has been commonly employed in the Anelastic Spherical Harmonics simulations \citep{miesch2000,brun2004}.
Although the sub-grid-scale physics in our model is different from that of these previous studies, height dependence of the diffusivities is fixed just for simplicity.

The sub-grid scale that we are considering in this study represents a scale smaller than that at which the magnetic energy excited by small-scale dynamos exceeds the kinetic energy of the turbulent flows (let us call this scale ``SSD scale'' hereafter).
We compute the convective motions on scales larger than SSD scale, assuming that the small-scale magnetic fields have little direct influence on these scales.
Typically, SSD scale is expected to lie between the inertial range of the turbulent flows \citep{hotta2015,hotta2016} so that the moderate Reynolds number is enough for this study.
It should be emphasized that we do not aim at any realistic solar magneto-convection simulations with this model but rather study the physical processes of convective velocity suppression under a simplifying assumption that the effects of small-scale magnetism can be modeled as sub-grid-scale diffusivities. 

\subsection{Radiative Energy Fluxes} \label{subsec:3-3}

In our model, the convectively unstable stratification is formed by injecting the artificial radiative energy flux $F_{*}$ from the bottom and extracting the same amount of energy flux through the upper boundary.
For specifying the radiative heating term, the functional form presented in \citet{featherstone2016} is adopted which can mimic the normalized radiative energy flux tabulated in a standard solar model.
The radiative heating $Q_{\mathrm{heat}}$ is assumed to be proportional to the background pressure,
\begin{eqnarray}
  Q_{\mathrm{heat}}(z)=\alpha (p_{0}(z)-p_{0}(z_{\mathrm{max}})),
\end{eqnarray}
where the normalization factor $\alpha$ is specified by the input energy flux $F_{*}$ as,
\begin{eqnarray}
  \alpha=F_{*}\left[\int_{0}^{z_{\mathrm{max}}} (p_{0}(z)-p_{0}(z_{\mathrm{max}})) dz \right]^{-1}.
\end{eqnarray}
The radiative energy flux $F_{\mathrm{r}}$ is then defined as,
\begin{eqnarray}
  F_{\mathrm{r}}(z)=F_{*}-\int_{0}^{z}Q_{\mathrm{heat}}(z') dz'.
\end{eqnarray}

The radiative cooling term at the surface $Q_{\mathrm{cool}}$ is given in a similar form to \citet{hotta2014} as,
\begin{eqnarray}
  Q_{\mathrm{cool}} &=& -\frac{\partial}{\partial z}F_{\mathrm{sf}}(z), \\
  F_{\mathrm{sf}}(z)&=& F_{*} \exp{\left[ -\left(\frac{z-z_{\mathrm{max}}}{H_{0}(z_{\mathrm{max}})} \right)^{2}\right]}.
\end{eqnarray}
Therefore, the cooling effect is localized near the top boundary and the thickness of the upper thermal boundary layer is mostly set by the local pressure scale height there.

\subsection{Numerical Method} \label{subsec:3-4}

\begin{table*}[]
  \begin{center} 
    \caption{Parameters used and calculated in our model}
    \begin{tabular}{cccccccccc} \hline\hline
      \renewcommand{\arraystretch}{1.8}
      Case & $\mathrm{Pr^{H}}$ &  $\mathrm{Pr^{V}}$ && $\mathrm{Re_{eff}}$ & $\Delta\langle s\rangle/\mathrm{M}_{*}^{2}$ & $\delta_{\mathrm{sub}}/\mathrm{M}_{*}^{2}$ & $z_{\delta = 0}/H_{\mathrm{r}}$ & $v_{\mathrm{rms}}/v_{*}$ & $E_{\mathrm{kin}}/(F_{*}H_{\mathrm{r}}^{3}/v_{*})$
      \renewcommand{\arraystretch}{1} \\ \hline
      H1V1.........  & $1$ & $1$ &:   & $38.09$ & $71.29$  & -2.49  & $0.74$ & $1.47$ & $67.79$ \\
      H2V2.........  & $2$ & $2$ &:   & $32.96$ & $80.23$  & -3.48  & $0.78$ & $1.32$ & $52.01$ \\ 
      H6V6.........  & $6$ & $6$ &:   & $28.70$ & $96.80$  & -5.83  & $0.90$ & $1.19$ & $40.41$ \\ 
      H1V6.........  & $1$ & $6$ &:   & $36.39$ & $104.22$ & -4.21  & $0.74$ & $1.44$ & $62.77$ \\
      H6V1.........  & $6$ & $1$ &:   & $30.64$ & $65.99$  & -3.55  & $0.80$ & $1.23$ & $44.98$ \\
      \hline
    \end{tabular}
  \end{center}
  \vspace{-1.3\baselineskip}
  \tablecomments{Horizontal and vertical Prandtl number $\mathrm{Pr^{H}}$ and $\mathrm{Pr^{V}}$ are the free parameters in our model. $\mathrm{Re_{eff}}$ is the volume- and time-averaged (nearly $40 \ H_{\mathrm{r}}/v_{*}$) effective Reynolds number defined by equation (\ref{eq:Re}). $\Delta \langle s \rangle=|\mathrm{max}\langle s \rangle-\mathrm{min}\langle s \rangle|$ measures the entropy difference across the numerical domain averaged over nearly $80 \ H_{\mathrm{r}}/v_{*}$ simulation time. $\delta_{\mathrm{sub}}$ denotes the minimum superadiabaticity that is usually achieved near the base. $z_{\delta =0}$ represents the depth where the superadiabaticity changes from negative to positive value. $v_{\mathrm{rms}}$ here denotes the volume-averaged rms velocity in each case. $E_{\mathrm{kin}}$ is the kinetic energy in a stationary state integrated over the entire volume and averaged over nearly $40 \ H_{\mathrm{r}}/v_{*}$ simulation time.} \label{table:1}
\end{table*}

We solve the equations (\ref{eq:conti})-(\ref{eq:ent}) numerically with periodic boundary conditions for horizontal directions (at $x = 0, \ x_{\mathrm{max}}$ and $y = 0, \ y_{\mathrm{max}}$) and impenetrable, stress-free boundary conditions for lower and upper boundaries ($z = 0,\ z_{\mathrm{max}}$).
The free boundary condition is applied for density and entropy, i.e., $\partial\rho_{1}/\partial z =0$, $\partial s/\partial z = 0$ at the top and the bottom.
The size of the numerical domain is set by $x_{\mathrm{max}} = y_{\mathrm{max}} = 8.72\ H_{\mathrm{r}}$ and $z_{\mathrm{max}} = 2.18 \ H_{\mathrm{r}}$.
The vertical domain spans $3$ density scale heights with $20$ density contrasts.

The magnitude of sub-grid-scale diffusivities are measured by the Reynolds number at the base $\mathrm{Re}_{*} \equiv v_{*}H_{\mathrm{r}}/\nu_{\mathrm{r}}$ and the Prandtl number which, in our simulations, is defined distinctly for both horizontally $\mathrm{Pr^{H}} \equiv \nu_{\mathrm{r}} /\kappa^{\mathrm{H}}_{\mathrm{r}}$ and vertically $\mathrm{Pr^{V}} \equiv \nu_{\mathrm{r}} /\kappa^{\mathrm{V}}_{\mathrm{r}}$.
Here, the velocity scale is estimated based on the input energy flux as $v_{*} \equiv (F_{*}/\rho_{\mathrm{r}} )^{1/3}$.
The energy flux $F_{*}$ is determined by specifying the modified Mach number $\mathrm{M}_{*}$ which is defined as $\mathrm{M}_{*} \equiv v_{*}/ \sqrt{p_{\mathrm{r}}/\rho_{\mathrm{r}}}$.
The values of the above dimensionless numbers for our simulations are given by $\mathrm{Re}_{*}=70$, $\mathrm{M}_{*}=1\times 10^{-2}$.
With this numerical setup, the subsequent amplitudes of the dimensionless entropy or the superadiabatic gradient would be on the order of $\mathcal{O}(M_{*}^{2})$.
The assumption of linearized equation of state is verified considering that $\rho_{1}/\rho_{\mathrm{r}}\approx p_{1}/p_{\mathrm{r}}\approx \mathrm{M}_{*}^{2}=10^{-4}$.
Prandtl numbers $\mathrm{Pr^{H,V}}$ are treated as free parameters in this study and given as listed in Table \ref{table:1}.
We should emphasize here that the kinetic viscosity $\nu$ is kept fixed in all simulations and we only change $\kappa^{\mathrm{H}}$ and $\kappa^{\mathrm{V}}$.
In fact, the SGS $\nu$ should also increase due to the Lorentz-force of the small-scale magnetic fields.
However, the dynamical effect originating from the change in $\nu$ is not considered and only the thermal effect is investigated in this study, which may under-estimate the degree of velocity suppression.

We use the fourth-order centered-differencing method for space and the fourth-order Runge-Kutta scheme for the time-integration \citep{vogler2005}.
The same artificial viscosity used in \citet{rempel2014} are implemented and added to velocity fields to stabilize numerical computations.
The code is parallelized using message passing interface (MPI).
For all our calculations, an uniform resolution of $288^{2} \times 72$ is used.
Each simulation is initiated by giving a small random perturbation for vertical velocity $v_{z}$ with the other variables $\rho_{1}$, $v_{x}$, $v_{y}$, and $s$ set to zero.

\section{Results} \label{sec:resultA} \label{sec:four}

\subsection{Overview} \label{subsec:4-1}

\begin{figure}[]
\figurenum{2}
\plotone{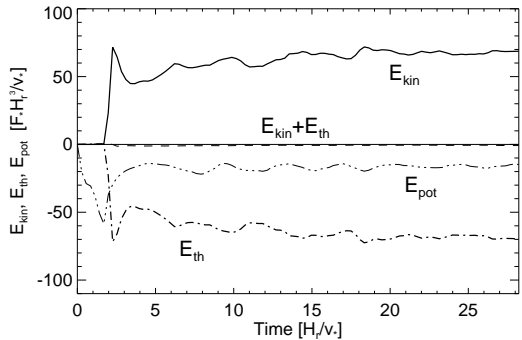}
\caption{Time series of volume-integrated kinetic energy (solid line), thermal energy (dot-dashed line), gravitational potential energy (dot-dot-dashed line), and the sum of thermal and kinetic energy (dashed line) for Case H1V1.} \label{fig:energyt}
\end{figure}

First, let us review the time evolution of the system.
In the following, we use the linear approximation for thermodynamic variables and also omit terms proportional to $\nabla\cdot (\rho_{0}\bm{v})$ that are small for low Mach number flows.
From equations (\ref{eq:motion}) and (\ref{eq:ent}), the energy equation is derived as,
\begin{eqnarray}
  \frac{\partial}{\partial t}\left(\frac{\rho_{0}}{2}\bm{v}^{2}+\rho_{0}T_{0}c_{\mathrm{v}}s \right)&=&-\nabla\cdot\left[\left(\frac{p_{0}}{\gamma -1}s+p_{1} \right)\bm{v} \right] \nonumber \\
  -\nabla\cdot\left(\frac{\rho_{0}}{2}|\bm{v}^{2}|\bm{v} \right)&+&\nabla\cdot(c_{\mathrm{v}}\rho_{0}T_{0}\bm{\kappa}\cdot\nabla s) \nonumber \\
  +\nabla\cdot(\bm{v}\cdot\bm{\Pi})&+&Q_{\mathrm{heat}}+Q_{\mathrm{cool}}.
\end{eqnarray}
Under the boundary conditions described earlier, a global energy conservation law can be expressed as,
\begin{eqnarray}
  \frac{\partial}{\partial t}\int_{V}\left(\frac{\rho_{0}}{2}\bm{v}^{2}+\rho_{0}T_{0}c_{\mathrm{v}}s \right)dV=0. \label{eq:cons}
\end{eqnarray}
Figure \ref{fig:energyt} shows the temporal evolution of total kinetic energy $E_{\mathrm{kin}}$, thermal energy $E_{\mathrm{th}}$, and gravitational potential energy $E_{\mathrm{pot}}$ in Case H1V1 for a reference.
The volume integrated energies are defined as,
\begin{eqnarray}
  E_{\mathrm{kin}}&\equiv& \int_{V} \frac{\rho_{0}}{2} \bm{v}^{2} \ dV, \\
  E_{\mathrm{th}}&\equiv& \int_{V} \rho_{0} T_{0} c_{\mathrm{v}} s \ dV, \\
  E_{\mathrm{pot}}&\equiv& \int_{V} \rho_{1}g H_{0} \ dV.
\end{eqnarray}
Note that, in our definition, the thermal energy $E_{\mathrm{th}}$ represents a perturbation with respect to the energy contained by the adiabatic background.
The internal energy of the system is expressed as $E_{\mathrm{int}}=E_{\mathrm{th}}+E_{\mathrm{pot}}$.
As prescribed by the equation (\ref{eq:cons}), a conservation of $E_{\mathrm{kin}}+E_{\mathrm{th}}$ can be well confirmed from Figure \ref{fig:energyt}.

Since the surface cooling can quickly produce highly superadiabatic stratification near the top boundary, the thermal convection typically sets in at an early stage $t \approx 1.5\ H_{\mathrm{r}}/v_{*}$ and a statistically stationary state is achieved after $t \ga 25\ H_{\mathrm{r}} /v_{*}$ as shown in Figure \ref{fig:energyt}.
We then further evolve the system for at least one thermal diffusion time $\tau_{\mathrm{diff}} \approx (\mathrm{Re}_{*} \mathrm{Pr^{V}}) H_{\mathrm{r}}/v_{*}$ after the statistically stationary state is reached.
From now on, we are going to discuss the results of this stage.
The statistical properties of the convection are investigated by temporally-averaging the data with a cadence of $0.9-1.2 \ H_{\mathrm{r}}/v_{*}$.

\subsection{Isotropic Thermal Diffusion} \label{subsec:4-2}

\begin{figure}[t]
\figurenum{3}
\plotone{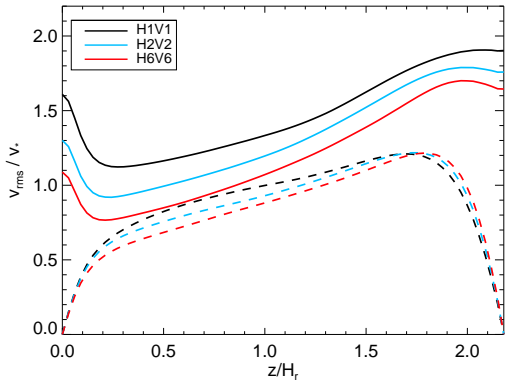}
\caption{The height dependence of the rms velocities for Case H1V1 (black), Case H2V2 (light-blue), and Case H6V6 (red). Solid and dashed lines represent total and vertical rms velocities, respectively. The profiles shown here have been averaged over nearly $40\ H_{\mathrm{r}}/v_{*}$ of simulation time.} \label{fig:vrms_iso126}
\end{figure}
\begin{figure}[t]
\figurenum{4}
\plotone{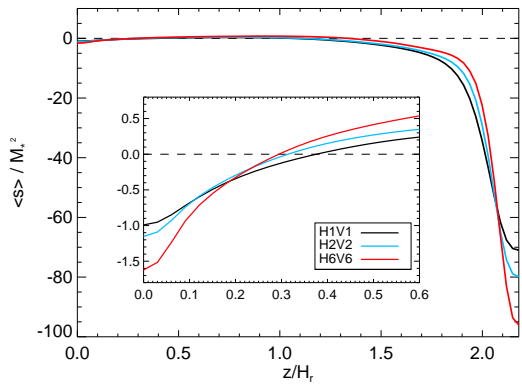}
\caption{Vertical profiles of entropy averaged over horizontal slices and time (spanning intervals of about $80\ H_{\mathrm{r}}/v_{*}$ late in the simulations) with the same Color-Case correspondence with Figure \ref{fig:vrms_iso126}. Inset: zoon-in the region near the bottom where the mean entropy profile is slightly subadiabatic.} \label{fig:ent_iso126}
\end{figure}

In this section, the results of Cases H1V1, H2V2, and H6V6 are discussed where thermal diffusion is isotropic ($\mathrm{Pr}=\mathrm{Pr^{H}}=\mathrm{Pr^{V}}$).
We try to make our argument that the change in the mean entropy stratification is mostly responsible for the convective velocity suppression in high-$\mathrm{Pr}$ regime by relating the suppressed convective velocities with the enhanced subadiabatic layers.
Figure \ref{fig:vrms_iso126} clearly shows that the rms velocity systematically declines as $\kappa$ is decreased ($\mathrm{Pr}$ is increased), as found by \citet{omara2016}.
The dashed lines in Figure \ref{fig:vrms_iso126} represent vertical rms velocities that are directly subject to buoyancy accelerations.
It is noteworthy that the vertical rms velocities are decreased in the bulk of the domain except for the top layer and correspondingly that the total rms velocities are suppressed mainly in the deeper layer.
The effective Reynolds number $\mathrm{Re_{eff}}$ is estimated in each case using the $z$-dependent rms velocities as,
\begin{eqnarray}
  \mathrm{Re_{eff}}=\frac{1}{z_{\mathrm{max}}} \int \frac{v_{\mathrm{rms}} H_{0}}{\nu}\ dz, \label{eq:Re}
\end{eqnarray}
and presented in the fourth column of Table \ref{table:1}.
In our model, the convection typically shows a moderate degree of turbulence with the typical value of $\mathrm{Re_{eff}} \approx 30$ which is of the same order as used in previous studies \citep{omara2016,kapyla2017b}.

Figure \ref{fig:ent_iso126} shows the mean entropy $\langle s \rangle$ normalized by $\mathrm{M}_{*}^{2}$ for each case.
Here, $\langle \ \rangle$ denotes the horizontal averaging.
In this paper, we normalize the entropy $s$ (and superadiabaticity $\delta$) by $\mathrm{M}_{*}^{2}$ to make a comparison with results with different input energy fluxes easier.
It is obvious from Figure \ref{fig:ent_iso126} that the highly-superadiabatic thermal boundary layer is formed near the top boundary, and the bulk of the domain is more close to adiabatic compared with the top layer.
Although the same form of surface cooling flux is imposed in all cases, the thickness of the upper thermal boundary layer $d_{\mathrm{t}}$ varies according to $\mathrm{Pr}$ due to the vertical thermal conduction near the surface.
As a result, the entropy difference between the top boundary and the bulk of the domain $\Delta \langle s \rangle \equiv|\mathrm{max}\langle s \rangle-\mathrm{min}\langle s \rangle| $ increases as $\mathrm{Pr}$ increases.
However, $\mathrm{Pr}$-dependence of the upper boundary layer is small compared with the previous studies \citep{omara2016}, having no scaling relations between the thermal diffusivity $\kappa$, the thickness of the thermal boundary layer $d_{\mathrm{t}}$, and the entropy difference $\Delta \langle s \rangle$, such as $d_{\mathrm{t}}\propto\kappa^{1/2}$ or $\Delta\langle s \rangle\propto \kappa^{-1/2}$ which can only hold for simulations imposing a diffusion-type upper boundary condition where the energy flux is released through the thermal conduction term \citep{featherstone2016}.

The zoom-in inset of Figure \ref{fig:ent_iso126} manifests firstly that the mean entropy tends to be slightly subadiabatic near the bottom boundary and secondly that the subadiabaticity there increases as $\mathrm{Pr}$ increases.
These results can be interpreted that the subadiabatic layer is formed by continuous depositions of low entropy transported by downflows.
When the thermal diffusivity $\kappa$ is reduced, the amount of low entropy fluids that can be retained by downflows during the descent becomes large, which leads to an enhancement of the subadiabaticity near the base.
The enhanced subadiabatic layer then suppresses downflow motions by alleviating the thermal contents of downflows and finally sets the net accumulation rate of the low entropy to the bottom.
In a statistically-stationary state, the continuous accumulation of low entropy is compensated by the vertical thermal conduction near the bottom boundary.

Some caution should be taken here regarding the impenetrable lower boundary condition in our idealized model.
It is true that this lower boundary condition indeed favorably affects the formation of this weakly-subadiabatic layer, helping the reflection and accumulation of low entropy around the base.
However, the existence of the slightly-subadiabatic convection zone has been repeatedly reported even in the numerical simulations of convective overshoots where a stably-stratified layer is included \citep{brummell2002,kapyla2017b} and also predicted by non-local semi-analytical model of solar overshoot layer \citep{xiong2001,rempel2004}.
Moreover, recent numerical simulations of solar overshoot region, which aim to achieve more realistic parameter regimes by imposing a much lower energy flux, estimated the depth of solar overshoot layer to be less than a few percent of the local pressure scale height there \citep{hotta2017}, suggesting that the bottom of the solar convection zone may act as an impenetrable wall to a good approximation.

\begin{figure}[t]
\figurenum{5}
\plotone{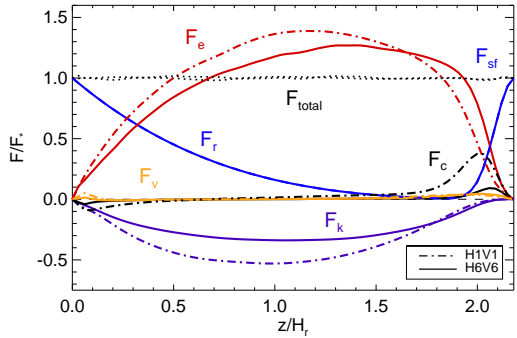}
\caption{Energy flux balance for Case H1V1 (dot-dashed lines) and Case H6V6 (solid lines). Shown on vertical axis are the energy fluxes normalized by injected energy flux $F_{*}$. The red, purple, yellow, and black lines represent the enthalpy flux $F_{\mathrm{e}}$, kinetic energy flux $F_{\mathrm{k}}$, viscous dissipative flux $F_{\mathrm{v}}$, and thermal conductive flux $F_{\mathrm{c}}$ calculated by the equations (\ref{eq:flux1})-(\ref{eq:flux4}) and averaged over time (nearly $80 \ H_{\mathrm{r}}/v_{*}$), respectively. The blue line shows a sum of time-independent radiative heating flux $F_{\mathrm{r}}$ and surface cooling flux $F_{\mathrm{sf}}$. The total energy flux $F_{\mathrm{total}}=F_{\mathrm{e}}+F_{\mathrm{k}}+F_{\mathrm{c}}+F_{\mathrm{v}}+F_{\mathrm{r}}+F_{\mathrm{sf}}$ is shown by black dotted line for both cases.} \label{fig:flux_iso16}
\end{figure}

\begin{figure*}[]
\figurenum{6}
\plotone{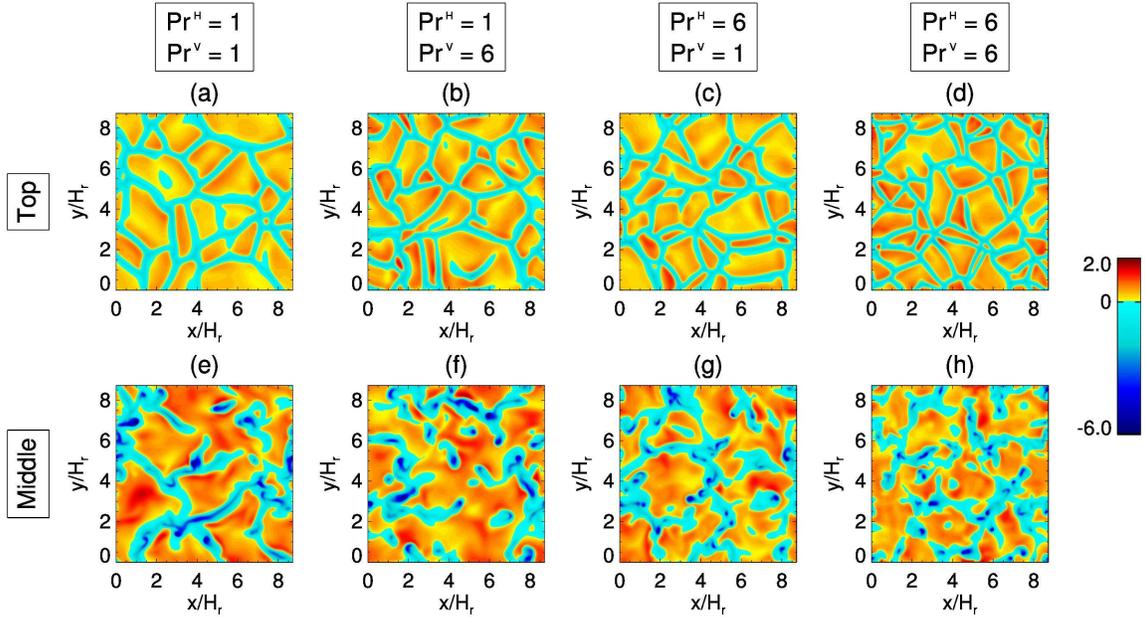}
\caption{Vertical velocity $v_{z}/v_{*}$ taken from statistically-stationary states for different values of $\mathrm{Pr^{H,V}}$: (a) and (e) for Case H1V1, (b) and (f) for Case H1V6, (c) and (g) for Case H6V1, and (d) and (h) for Case H6V6. Upper and lower panels show the horizontal section near the surface $z/H_{\mathrm{r}}=2.1$ and middle of the convection zone $z/H_{\mathrm{r}}=1.0$, respectively.} \label{fig:vz}
\end{figure*}
\begin{figure*}[]
\figurenum{7}
\plotone{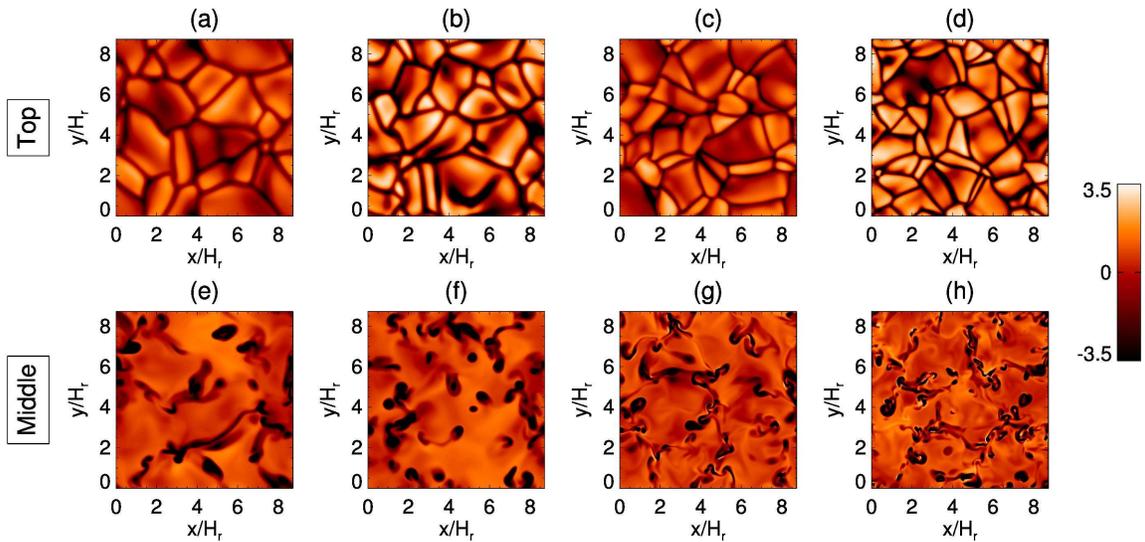}
\caption{Entropy perturbations $\rho_{0}(s-\langle s \rangle)/(\rho_{\mathrm{r}} \mathrm{M}_{*}^{2})$ with the same configuration as Figure \ref{fig:vz}.} \label{fig:ent}
\end{figure*}

\begin{figure*}[]
\figurenum{8}
\plotone{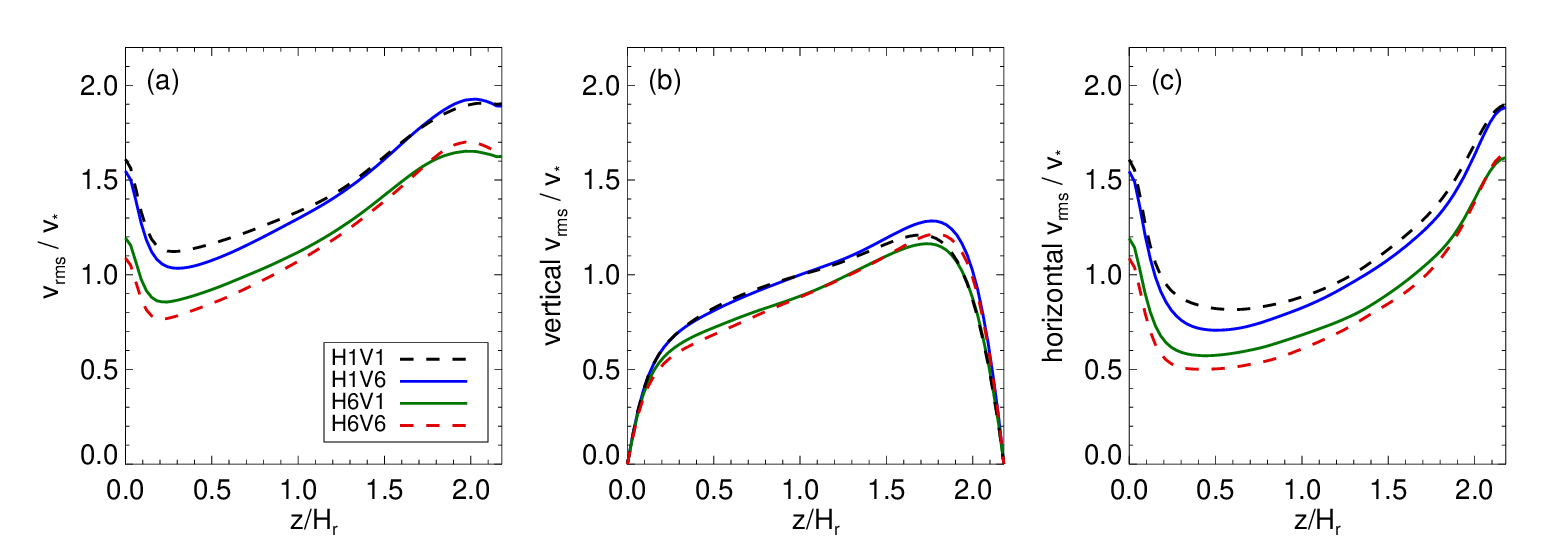}
\caption{Height dependence of rms velocities normalized by $v_{*}$ for Case H1V1 (black dashed), Case H1V6 (blue solid), Case H6V1 (green solid), and Case H6V6 (red dashed). Panels (a), (b), and (c) show the total, vertical, and horizontal rms velocities averaged over nearly $40 \ H_{\mathrm{r}}/v_{*}$ simulation time, respectively.} \label{fig:vrms_an6}
\end{figure*}

Figure \ref{fig:flux_iso16} shows horizontally-averaged energy fluxes across the numerical domain for Case H1V1 and H6V6.
The definitions of the enthalpy flux $F_{\mathrm{e}}$, the kinetic energy flux $F_{\mathrm{k}}$, the thermal conductive flux $F_{\mathrm{c}}$, and the viscous dissipative flux $F_{\mathrm{v}}$ are,
\begin{eqnarray}
  F_{\mathrm{e}}&=&\frac{p_{0}}{\gamma -1} \langle s_{\ } v_{z} \rangle + \langle p_{1} v_{z} \rangle, \label{eq:flux1}  \\
  F_{\mathrm{k}}&=&\frac{\rho_{0}}{2}\langle \bm{v}^{2} v_{z}\rangle, \\
  F_{\mathrm{c}}&=&-\kappa^{\mathrm{V}} \frac{p_{0}}{\gamma -1}\frac{\partial \langle s \rangle}{\partial z}, \\
  F_{\mathrm{v}}&=&-\langle \bm{v}\cdot \bm{\Pi} \rangle. \label{eq:flux4}
\end{eqnarray}
Due to the velocity suppression in Case H6V6, both the amplitudes of enthalpy flux and kinetic energy flux decrease.
The reduction in the vertical thermal conductive flux near the surface in Case H6V6 is compensated by the vertically-upward peak shift of the enthalpy flux.

It should be noted that the enthalpy flux takes positive value throughout the numerical domain so that the thermal energy is transported vertically upward even near the base where the mean stratification is subadiabatic.
Therefore, this subadiabatic layer formed near the base is not an overshooting layer where downflows are quickly decelerated by buoyancy and thus the enthalpy flux becomes negative.
Rather it should be regarded as a result of non-local heat transport of downflow plumes, which cannot be described by the typical local mixing models assuming the enthalpy flux proportional to the local superadiabaticity.
Recently, \citet{brandenburg2016} modified the mixing-length theory incorporating the effects of non-local heat transport by cold downflow plumes and showed that the enthalpy flux can be upward even in the subadiabatic region.
We will discuss on this issue later in Section \ref{subsec:5-1}.
Even though the subadiabaticity near the base is not strong enough to reverse the sign of enthalpy flux to be a overshooting layer, this layer has significant effects on the convective velocity amplitudes.

\subsection{Anisotropic Thermal Diffusion} \label{subsec:4-3}

Next, we examine results of Case H1V6 and Case H6V1 where vertical and horizontal thermal diffusivities $\kappa^{\mathrm{V}}$ and $\kappa^{\mathrm{H}}$ are decreased independently so that several thermal effects can be separated.
In Case H1V6, we drop only $\kappa^{\mathrm{V}}$ to the same value used in Case H6V6 while keeping $\kappa^{\mathrm{H}}$ identical to Case H1V1.
In Case H6V1, on the other hand, only $\kappa^{\mathrm{H}}$ is dropped to the same level of Case H6V6 with $\kappa^{\mathrm{V}}$ unchanged from Case H1V1.

First, the overall convection patterns are explained.
Figure \ref{fig:vz} and Figure \ref{fig:ent} show vertical velocities and entropy perturbations for Cases H1V1, H1V6, H6V1, and H6V6 from left to right panels with upper and lower panels presenting the horizontal cuts near the top and at the middle, respectively.
In Figure \ref{fig:vz}, $\mathrm{Pr}$-dependency of the vertical convection is more apparent in the horizontal cuts at the middle layer (panels (e)-(h)):
We can see that downflows become more pointwise when $\kappa^{\mathrm{H}}$ is decreased.
In other words, convective structure is qualitatively changed from lane-type downflows to plume-type downflows as $\kappa^{\mathrm{H}}$ decreases.
Moreover, it is also obvious that the overall amplitudes of upflows and downflows are reduced in Cases H6V1 and H6V6 where $\kappa^{\mathrm{H}}$ is reduced, which will be confirmed later in Figure \ref{fig:vrms_an6}.

As for entropy, it is clear from Figure \ref{fig:ent}(a)-(d) that the amplitude of the entropy perturbation at the surface becomes large for Cases H1V6 and H6V6.
This is because the suppression in $\kappa^{\mathrm{V}}$ makes the upper thermal boundary layer steeper, leading to a thinner highly-superadiabatic surface layer and an enlargement of the absolute value of entropy difference.
The change in $\kappa^{\mathrm{H}}$, on the other hand, only plays a role in forming thinner downflow lanes near the surface.
As we go into deeper convection zone, however, the entropy distributions are dominantly characterized by $\kappa^{\mathrm{H}}$, as shown in Figure \ref{fig:ent}(e)-(h):
Downflow plumes are able to retain their low entropy at small scales in Cases H6V1 and H6V6 where horizontal diffusion is highly prohibited.
In Cases H1V1 and H1V6, in contrast, efficient horizontal heat exchanges blur the entropy at deeper convection zone.

\begin{figure*}[]
\figurenum{9}
\plotone{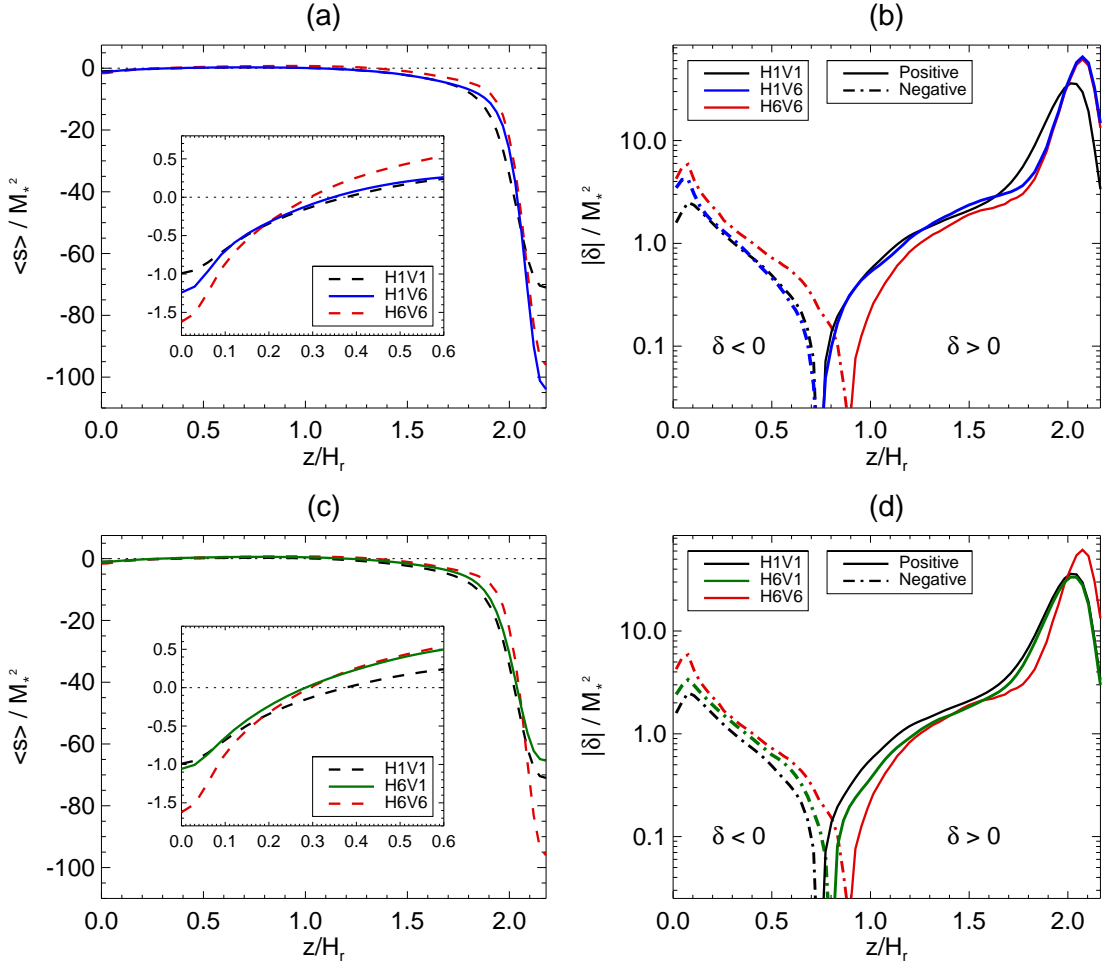}
\caption{Horizontally-averaged entropy profiles and the corresponding superadiabaticity $\delta=\nabla-\nabla_{\mathrm{ad}}$ profiles. Upper and lower panels show the results of Case H1V6 and Case H6V1, respectively. The results of Case H1V1 and H6V6 are also shown for comparison in every panel. In panels (a) and (c), mean entropy profiles $\langle s\rangle$ normalized by $\mathrm{M}_{*}^{2}$ are plotted for Case H1V1 (black dashed), Case H1V6 (blue solid), Case H6V1 (green solid), and Case H6V6 (red dashed) with the zoom-in inset focusing on the region near the bottom. In panels (b) and (d), the superadiabaticity $\delta$ also normalized by $\mathrm{M}_{*}^{2}$ are plotted with the same Case-Color correspondence. The absolute values of the superadiabaticity are shown on a vertically-logarithmic scales, with the negative values shown by dot-dashed lines and positive values by solid lines. The profiles shown here have been averaged over time (spanning intervals of $80-100 \ H_{\mathrm{r}}/v_{*}$ late in the simulations).} \label{fig:entdel_an45}
\end{figure*}
\begin{figure}[]
\figurenum{10}
\plotone{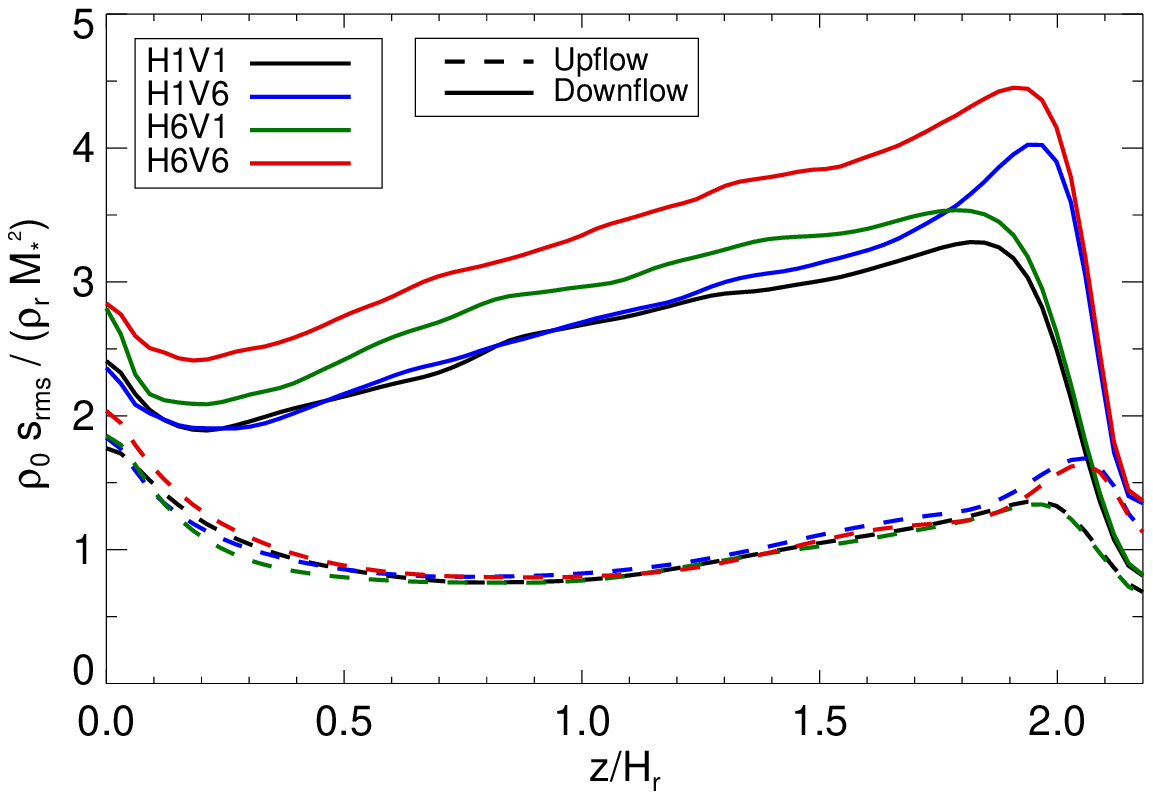}
\caption{The rms values of entropy perturbation averaged over time (nearly $40 \ H_{\mathrm{r}}/v_{*}$) and normalized by background density $\rho_{0} s_{\mathrm{rms}}$ against height. The mean is taken each for upflows and downflows and the calculated rms values are shown by dashed and solid lines, respectively. The same Case-Color correspondence as Figure \ref{fig:entdel_an45} is used. } \label{fig:entrms}
\end{figure}

Figure \ref{fig:vrms_an6} shows distributions of total, vertical, and horizontal rms velocities for Cases H1V1, H1V6, H6V1, and H6V6.
We can confirm from Figure \ref{fig:vrms_an6}(a) that the significant reduction of rms velocity only occurs in Case H6V1 and H6V6 in which $\kappa^{\mathrm{H}}$ is decreased (see also the volume-averaged rms velocity values tabulated in the eighth column of Table \ref{table:1}).
The volume-integrated kinetic energies $E_{\mathrm{kin}}$ are calculated and presented in the ninth column of Table \ref{table:1}.
$E_{\mathrm{kin}}$ is highly reduced up to $59.6 \%$ for Case H6V6 and $66.3 \%$ for Case H6V1 with respect to that of Case H1V1, whereas in Case H1V6 the kinetic energy is suppressed only slightly ($92.6 \%$).
Especially, it is obvious from Figure \ref{fig:vrms_an6}(b) that the values of vertical rms velocities are mostly determined by $\kappa^{\mathrm{H}}$, except for the top surface where $\kappa^{\mathrm{V}}$ influences the convective properties by increasing the Rayleigh number $\mathrm{Ra}$.
Therefore, it is concluded here that the prohibition of horizontal, not vertical, diffusive transport of entropy between warm upflows and cold downflows is responsible for the velocity suppression in high-$\mathrm{Pr}$ regime.
We attribute this convective velocity suppression to the change of mean entropy profile as follows.

Figure \ref{fig:entdel_an45} shows profiles of the mean entropy and corresponding superadiabaticity calculated by,
\begin{eqnarray}
  \delta &=& \nabla -\nabla_{\mathrm{ad}} \nonumber \\
         &=& -\frac{H_{0}}{\gamma}\frac{\partial \langle s \rangle}{\partial z},
\end{eqnarray}
for Case H1V6 and H6V1 at upper and lower panels, respectively.
The amount of entropy difference between the top and the bulk $\Delta \langle s \rangle$ is substantially influenced by $\kappa^{\mathrm{V}}$ as shown in Figure \ref{fig:entdel_an45}(a) and (c) (see also the fifth column of Table \ref{table:1}).
For example, in Case H1V1 (black) and Case H6V1 (green) where $\mathrm{Pr^{V}}=1$ and $\kappa^{\mathrm{V}}$ is relatively large, strong vertical thermal diffusion near the top efficiently alleviates the highly-superadiabatic entropy gradient, resulting in the small entropy difference $\Delta \langle s \rangle$.

Figure \ref{fig:entdel_an45}(b) shows that the superadiabaticity profile is almost unaffected by a decrease in $\kappa^{\mathrm{V}}$ in the bulk of the domain.
On the other hand, we can clearly observe from Figure \ref{fig:entdel_an45}(d) that the subadiabatic layer is enhanced and extended vertically upward in the lower portion of the domain and also that the mean stratification in the bulk becomes less superadiabatic from Case H1V1 (black) to Cases H6V1 (green) and H6V6 (red) as $\kappa^{\mathrm{H}}$ decreases.
Thus, $\delta$ is only sensitive to $\kappa^{\mathrm{H}}$ and this is why huge reductions of convective velocity occur in Cases H6V1 and H6V6.
Both the enhancement of subadiabaticity and weakening of superadiabaticity lead to the reduction of the net buoyancy acceleration and suppression of the convective amplitudes.

The different sensitivity of $\delta$ on $\kappa^{\mathrm{V}}$ and on $\kappa^{\mathrm{H}}$ comes from the fact that downflows can retain their cold entropy only when the horizontal thermal diffusion is inhibited, as clearly shown in Figure \ref{fig:entrms} where entropy fluctuations $s_{\mathrm{rms}}$ of upflows and downflows are plotted.
Firstly, it is obvious that only the entropy contents of downflows are modified and the thermal properties of upflows are almost unchanged by $\kappa^{\mathrm{H,V}}$.
In Case H1V6 (blue), thermal fluctuation of downflows is mostly identical to that of Case H1V1 (black) in the bulk, which means that downflows tend to quickly lose their cold entropy via the horizontal thermal diffusion even though strong entropy deficits are generated at the top.
In Case H6V1 (green), on the other hand, downflows can retain their entropy deficits against horizontal thermal diffusion so that they can greatly contribute to the overall accumulations of low entropy around the base.
It is considered that the amount of negative entropy perturbation that can be transported by downflows and accumulated near the base should largely set the subadiabaticity there.

In fact, these results should be interpreted with some caution because the thermal diffusivities have a vertical dependence expressed by the equation (\ref{eq:v-depend}). 
In principle, the profile of $\delta$ in a stationary state should be determined by both  $\kappa^{\mathrm{H}}$ and $\kappa^{\mathrm{V}}$; the former one adjusts the amount of low entropy that is supplied to the base and the latter one sets the final entropy profile after the vertical thermal relaxation. 
Although the $\kappa^{\mathrm{H}}$-dependence of $\delta$ near the base may be emphasized in our model by originally prohibiting the vertical thermal conduction in the lower part in all cases, our conclusion that the decrease in $\kappa^{\mathrm{H}}$ is crucial for the enhancement of subadiabatic layer and for the convective velocity suppression is indeed supported by results of another set of simulations where spatially-uniform $\kappa^{\mathrm{H,V}}$ is used (not shown here):
The suppression in $v_{\mathrm{rms}}$ occurs only when $\kappa^{\mathrm{H}}$ are decreased.
When the amplitude of the entropy perturbation is increased, low entropy material is transported more to the base.
This process leads to an extension of the subadiabatic layer to the upper part of the convection zone and thus makes the bulk stratification more subadiabatic and less superadiabatic.
On the other hand, effects of the vertical thermal diffusion are restricted to the lower and upper thermal boundary layers and does not largely affect the mean entropy stratification in the bulk which is more close to adiabatic.
As a result, $v_{\mathrm{rms}}$ is almost unaffected by $\kappa^{\mathrm{V}}$.

\begin{figure}[]
\figurenum{11}
\plotone{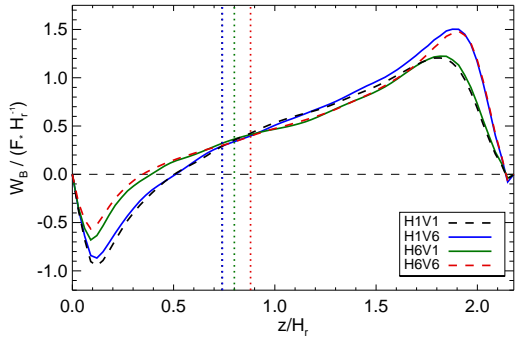}
\caption{Work densities done by buoyancy forces averaged over horizontal slices and time ($40\ H_{\mathrm{r}}/v_{*}$) for Case H1V1 (black dashed), Case H1V6 (blue solid), Case H6V1 (green solid), and Case H6V6 (red dashed). The vertical dotted lines denote the heights from which the mean stratification becomes superadiabatic ($z_{\delta =0}$) with the same Case-Color correspondence.} \label{fig:Wbuo}
\end{figure}

In order to quantitatively examine the effects of the enhanced subadiabatic layer, work density done by buoyancy force is calculated as,
\begin{eqnarray}
  W_{\mathrm{B}}=-\frac{\partial p_{1}}{\partial z}v_{z}-\rho_{1}g v_{z},
\end{eqnarray}
and presented in Figure \ref{fig:Wbuo}.
The buoyancy work becomes positive in the upper convection zone and negative in the lower convection zone.
This general tendency is consistent with the fact that superadiabatic (subadiabatic) stratification accelerates (decelerates) thermal convection.
The existence of the region near the bottom where the buoyancy work is negative $W_{\mathrm{B}}<0$ can be thus regarded as a clear evidence of convection suppression by the subadiabatic layer.
In Case H6V1 (green) and Case H6V6 (red), the positive buoyancy work is reduced due to the less superadiabatic stratification in the upper convection zone except for the surface region.
Near the bottom, on the other hand, the absolute values of negative buoyancy work is decreased in these two cases where subadiabaticity is enhanced, which at first glance seems contradictory to our argument.
We consider that the buoyancy force can do less negative work (to decelerate convection) just because the vertical velocity amplitudes are reduced in these two cases.
Note here that the buoyancy works calculated by the above equation just represent the energy conversion rate between thermal and vertical kinetic energies in a statistically stationary state.
A large amount of positive buoyancy work does not necessarily result in a large amount of kinetic energy or large convective amplitudes.

Vertical dotted lines shown in Figure \ref{fig:Wbuo} denote the heights from which mean stratifications change from subadiabatic to superadiabatic $z_{\delta =0}$ (see the seventh column of Table \ref{table:1}).
Although the subadiabatic stratification eventually results in the negative buoyancy work $W_{\mathrm{B}}<0$ in the lower convection zone, the deceleration actually sets in below the critical height $z_{\delta =0}$ so that there is a gap zone in-between where the mean stratification is subadiabatic $\delta <0$ but buoyancy work is still positive $W_{\mathrm{B}}>0$.
The origin of this nonlocalness comes from the fact that the thermal fluctuations possessed by downflows (larger density than surroundings for instance) must be gradually modified as downflows travel inside the subadiabatic zone and become less heavy \citep{hotta2017}.
Figure \ref{fig:Wbuo} shows that the downflows are quickly decelerated after penetrating into the subadiabatic zone for Cases H1V1 (black) and H1V6 (blue) where the thermal contents of downflows are relatively small as shown in Figure \ref{fig:entrms}.
However, the nonlocalness become more substantial as downflows contain colder entropy from Case H1V1 (black) to Case H6V1 (green) to Case H6V6 (red).
This is because the amount of thermal fluctuations necessary for reversing the sign of buoyancy work gets larger in Case H6V1 and H6V6, so that it becomes more difficult and needs longer distance to achieve $W_{\mathrm{B}}<0$ for each downflow.
Therefore, the deceleration region is localized near the bottom in Cases H6V1 and H6V6, and as a result, it helps for downflows to reach the base and to reinforce the subadiabaticity by accumulating the low entropy there.
It should be emphasized again that since the subadiabaticity is on the order of the local Mach number square $\delta_{\mathrm{sub}}\approx \mathcal{O}(\mathrm{M}_{*}^{2})$ (see also sixth column of Table \ref{table:1}) and not strong enough to be an overshoot layer, downflows only feel gradual decelerations and thus the buoyancy works $W_{\mathrm{B}}<0$ are not simultaneously responded.

In summary, we have shown up to here by comparing the results of Cases H1V1, H1V6, H6V1, and H6V6 that the suppression of $\kappa^{\mathrm{H}}$ is essential in establishing the enhanced and extended subadiabatic layer near the base and in decreasing the whole convective amplitudes.
All of these results presented in this section are in qualitative accordance with the velocity suppression mechanism discussed in Section \ref{sec:two}.

\section{DISCUSSION} \label{sec:five}

\subsection{Implications for Solar Convection} \label{subsec:5-1}

In the preceding sections, we have explained how the convective flow speed could be suppressed when the horizontal thermal diffusivity $\kappa^{\mathrm{H}}$ is decreased.
Since the main focus of our research is to investigate the physical mechanism of velocity suppression and we do not aim at any realistic solar convection modeling, some caution is needed when discussing the applicability of our results to the solar convective conundrum.
Huge discrepancy of convective amplitudes (by more than two orders of magnitude) between the helioseismic finding of \citet{hanasoge2012} and the numerical simulation of \citet{miesch2008} was found mainly on the low wavenumber subsurface flow (with the spherical harmonic degree $l\approx 10$), which in other words suggests that the giant cells may not exist or their amplitudes may be far smaller than previously considered.
Considering that our numerical simulations employ a simplified Cartesian box with periodic boundaries in horizontal directions, discussions on the global-scale ($l\approx 10$) flow amplitudes are beyond the scope of this paper.

Nonetheless, our results have profound implications on the solar convective structure.
In general, it is considered that the excess in the low wavenumber power may reflect too large convective amplitudes in a deeper convection zone \citep{lord2014}.
A main advantageous feature of our proposed thermal effect is that convective amplitudes in a deeper layer can be selectively suppressed along the enhancement of the subadiabaticity there.
This may potentially alleviate the convective conundrum by suppressing the deep convection and decreasing the low wavenumber power of the subsurface horizontal flow.

\begin{figure}[]
\figurenum{12}
\plotone{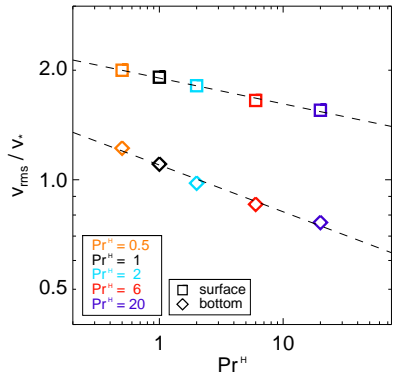}
\caption{The local-maximum rms velocities near the surface (by square points) and local-minimum rms velocities near the bottom (by diamonds) on double-logarithmic scale for different values of $\mathrm{Pr^{H}}$. Each dashed line denotes the fit to the results of diamonds and square points, showing a power-law functions with index 0.071 and 0.129, respectively.} \label{fig:scaling}
\end{figure}

Then, how much convective velocities can be reduced at most via the thermal effect?
Can the thermal effect be efficient enough to fully resolve the discrepancies between observations and simulations?
To derive some hints for answering these questions, another set of numerical simulations is conducted where only the horizontal Prandtl number $\mathrm{Pr^{H}}$ is systematically increased from $0.5$ up to $20$ with the vertical Prandtl number fixed ($\mathrm{Pr^{\mathrm{V}}}=1$).
The numerical domain is horizontally restricted to $[ 0 < x_{\mathrm{max}}, \ y_{\mathrm{max}} < 6.54 \ H_{\mathrm{r}}]$ and an uniform resolution of $216^{2}\times 72$ is used except for $\mathrm{Pr^{H}=20}$ Case where a resolution is increased up to $648^{2}\times 288$ in order to keep the influence of the numerical diffusivity sufficiently small.

Figure \ref{fig:scaling} shows the results of rms velocities $v_{\mathrm{rms}}$ obtained for different values of $\mathrm{Pr^{H}}$.
Local-maximum (local-minimum) values of the rms velocities near the surface (bottom) are plotted on a double-logarithmic scale.
The thermal effect does not saturate within the parameter regime investigated where Prandtl numbers are kept still moderate ($\mathrm{Pr^{H}} \le 20$) so that the rms velocities are found to monotonically decrease as $\mathrm{Pr^{H}}$ increases.
Figure \ref{fig:scaling} confirms that the rms velocities are more suppressed near the base than in the surface region.
The scaling relations of maximum rms velocity near the surface and minimum rms velocity near the bottom are calculated as,
\begin{eqnarray}
  v_{\mathrm{rms}}\propto \left\{
  \begin{array}{ll}
    \mathrm{Pr^{H}}^{-0.071} & \ (\mathrm{near \ the \ surface}) \\
    \mathrm{Pr^{H}}^{-0.129} & \ (\mathrm{near \ the \ bottom}).
  \end{array}
  \right. \label{eq:scaling}
\end{eqnarray}
Although this result shows a promising feature, it infers that the $\mathrm{Pr^{H}}$-dependence of our proposed thermal effect is relatively weak and that a significantly large effective Prandtl number on the order of $10^{7}$ is required to lower the velocity amplitude by $1/10$.

Several careful considerations are needed especially when trying to apply this result to the Sun or to the other numerical systems.
First of all, we remind the reader that the SGS viscous diffusivity $\nu$ (and thus the Reynolds number $\mathrm{Re}_{*}$) is fixed in all calculations for simplicity.
We must note that the scaling relation above is derived for the parameter regime employed in our numerical setup and that the scaling index may vary according to $\mathrm{Re}_{*}$.
For example, since we impose large viscous and thermal SGS diffusivities with the typical Reynolds number $\mathrm{Re_{eff}}$ calculated as $30-40$, the parameter regime studied in our simulations is laminar.
It is expected that, if SGS $\nu$ is decreased and the Reynolds number $\mathrm{Re}$ (and thus Rayleigh number $\mathrm{Ra}$) increases, the thermal effect would become ineffective and $v_{\mathrm{rms}}$ would cease to diminish being independent of the values of diffusivities \citep{featherstone2016}.

Secondly, determining whether or not the thermal effect saturates for higher $\mathrm{Pr}$, or if do, determining the saturated value of $v_{\mathrm{rms}}$ or the critical $\mathrm{Pr}$ would be another important issue that needs to be investigated when discussing the applicability of the scaling relation (equation (\ref{eq:scaling})).
Note that in the previous numerical study of \citet{omara2016} the saturation and limitation of velocity suppression in high-$\mathrm{Pr}$ regime was mainly discussed in relation to the thickness of the upper thermal boundary layer which scales as $d_{t}\propto \kappa^{1/2}$ owing to their conductive-type upper boundary condition: \citet{omara2016} argued that their suppression mechanism would become unphysical when $\kappa$ is decreased up to the point where $d_{t}$ approaches to the actual depth of the photospheric boundary of the Sun.
In our model, on the other hand, the depth of the upper boundary layer is set by the artificial surface cooling function and is almost independent of $\kappa$.
Therefore, the thermal effect discussed in this paper should not saturate in the same way as the model of \citet{omara2016}.
Instead, it is expected that in our model the saturation occurs when the thermal diffusion becomes negligible and accumulations of low entropy by downflows becomes unable to further change the mean entropy stratification:
If $\kappa$ becomes small enough to make the vertical thermal conduction essentially ineffective, the subadiabaticity near the base is determined so that the strong buoyant decelerations can stop downflows before they reach the bottom boundary and limit their supply of low entropy to the base.

From another side, the mean superadiabatic stratification of the solar interior has been estimated based on the mixing-length theory.
Although a great agreement between mixing-length models and solar surface convection simulations makes this model a helpful tool to describe the solar convection \citep[e.g.,][]{trampedach2011}, its reliability in the deep convection zone is still elusive:
In fact, mixing-length models typically predict the existence of giant convective cells in the deep convection zone whose signals can hardly be captured by helioseismology \citep{hanasoge2012}.
As already pointed out in Section \ref{subsec:4-2}, the subadiabatic layer formed in the lower part of our numerical domain reflects the nonlocal effect of heat transport.
The importance of the nonlocal treatment of mixing-length model has been repeatedly recognized mainly in the context of solar overshoot modeling.
The nonlocal convective overshoot models naturally predict an extended weakly-subadiabatic layer above the radiation zone \citep{xiong2001,rempel2004}, which can typically extend up to $r/R_{\odot}=0.75 - 0.8$ depending the nonlocality of convection \citep{skaley1991}.
Recently, \citet{brandenburg2016} modified the stellar mixing-length expression of the enthalpy flux by considering an additional term to incorporate the effects of nonlocal heat transport by strong downflow plumes, which enables even a weakly-subadiabatic region to transport the enthalpy upward.
The importance of our study lies in that the effect of the enhanced subadiabatic layer on convection is connected for the first time to the change in the effective Prandtl number via the strongly nonlocal energy transport by convective downflows which has been widely recognized since \citet{spruit1997}.
Our results offer a possibility that, if downflows can retain their low entropy in effectively high-$\mathrm{Pr}$ regime and the nonlocality accordingly increases, the subadiabatic lower convection zone would be enhanced and extended upward in the lower part of the convection zone.

\subsection{Possible Effects of Rotation} \label{subsec:5-2}

While we discussed so far the thermal effects on the convective velocity suppression in a non-rotating system for simplicity, it would be instructive to address some possible rotational effects on the proposed suppression mechanism.
In general, Coriolis force tends to bend downflows into a longitudinal direction.
It is thus expected that the subadiabatic layer near the bottom become more difficult to be formed due to the inefficient transport of low entropy by downflows.
If the rotational effects are too strong ($\mathrm{Ro} \ll 1$), the $\mathrm{Pr}$-dependency of convective amplitudes is expected to diminish, because downflows are quickly distorted and the convectional structure should be dominated by the Taylor-Proudman state being independent of the thermal properties.
Nonetheless, the proposed thermal effects may provide some important implications on the solar convection zone dynamics if we focus on a regime where rotational effects are moderate.

If the rotational effects are relatively weak ($\mathrm{Ro} \ga 1$), the subadiabaticity at the bottom is expected to exhibit substantial latitudinal dependence since Coriolis forces can work effectively on downflows near the equatorial region whereas downflows near the polar region barely feel the rotational effects.
As a result, a highly-subadiabatic layer should be formed near the pole and less subadiabatic (or even superadiabatic) layer should be formed near the equator, leading to a positive latitudinal gradient of superadiabaticity $\partial \delta/\partial \theta>0$.
This has a considerable influence on the thermal wind balance of the solar differential rotation.
It has been argued that the negative latitudinal entropy gradient is needed for explaining the non-cylindrical rotational profile of the observed differential rotation via the thermal wind balance \citep[e.g.,][]{miesch2005review}.
Although there are several theoretical explanations on the origin of this latitudinal entropy gradient \citep[e.g.,][]{kitchatinov1995,rempel2005,masada2011}, the global convection simulations can hardly reproduce it in a self-consistent manner and thus an ad-hoc latitudinal entropy variation has been commonly imposed at the bottom boundary in order to artificially break the Taylor-Proudman constraint \citep{miesch2006,fan2014}.
The latitudinal dependence of the superadiabaticity in the lower convection zone as a natural consequence of the proposed thermal effects in high-$\mathrm{Pr}$ regime may offer a promising mechanism to create the latitudinal entropy gradient at the base in a self-consistent manner via the interaction with the clock-wise (anti clock-wise) meridional circulation in the southern (northern) hemisphere \citep{rempel2005}.

Another concern related to the rotational effects would be the angular momentum transport by the turbulent Reynolds stress \citep{kitchatinov1993}.
This issue is not only essential for our understanding of differential rotation but also for determining the meridional circulation structure which, despite its importance for flux-transport dynamo model, varies significantly depending on theoretical models of the Reynolds stresses \citep{bekki2017}, global parameters employed in 3D simulations \citep{passos2015,featherstone2015}, and inversion techniques of the helioseismic observations \citep{zhao2013,rajaguru2015}.
As already seen in Section \ref{subsec:4-2}, the convectional structure in high-$\mathrm{Pr}$ regime is qualitatively different from those in $\mathrm{Pr}\approx 1$ which has been commonly investigated in great detail: High-$\mathrm{Pr}$ thermal convection is characterized by strong downflowing plumes descending deeper into convection zone across several pressure scale heights, transporting highly-concentrated cold entropy in a non-local way.
It would be non-trivial, therefore, whether we can apply the knowledge on the turbulent Reynolds stress derived from $\mathrm{Pr}\approx 1$ calculations to the solar convection zone whose effective Prandtl number might be larger than unity.
Further numerical work is needed in order to investigate how the properties of the turbulent momentum transport depend upon the Prandtl number $\mathrm{Pr}$ as well as on the Rossby number $\mathrm{Ro}$. 

\section{SUMMARY AND CONCLUSIONS} \label{sec:six}

In this paper, we have investigated one possible velocity suppression mechanism, motivated by the recently-recognized problem that the current solar convection simulations may be over-estimating the amplitudes of deep convection \citep[e.g.,][]{hanasoge2016review}.
We have especially focused on one interesting feature of the small-scale magnetism that it may decrease the effective thermal diffusivity $\kappa$ by inhibiting the small-scale turbulent mixing of entropy between warm upflows and cold downflows \citep{hotta2015}.
By conducting a set of thermal convection simulations where the effects of small-scale magnetism are incorporated as SGS diffusivities, we have shown in Section \ref{subsec:4-2} that the convective velocities are suppressed as we decrease the effective thermal diffusivity $\kappa$ through the enhancement of the subadiabatic layer which is formed near the base of the convection zone.

We further introduced the anisotropy of thermal diffusion in Section \ref{subsec:4-3} to finally conclude that the decrease in the horizontal thermal diffusivity $\kappa^{\mathrm{H}}$ is critical for the convective velocity suppression, which is consistent with the results of small-scale dynamo simulations \citep{hotta2015}.
This can be interpreted that the strong inhibition of horizontal entropy diffusion can promote the efficient transport of low entropy, and hence, promote the enrichment of the subadiabatic layer formation.

These results are synthesized to yield a picture of how the deep solar convection may be suppressed, which we outline as follows.
\begin{enumerate}
  \renewcommand{\labelenumi}{(\Roman{enumi}).}
\item 
  If small-scale dynamos are fully excited within the solar convection zone, convection is essentially magnetized and can operate in an effectively high-$\mathrm{Pr}$ regime \citep{hotta2015,hotta2016}.
\end{enumerate}
  This is because the Lorentz-force of small-scale magnetism may act as an effective viscosity, and as a result, the effective Reynolds number $\mathrm{Re_{eff}}$ for the flows on scales larger than the energy containing scale of the magnetic field is decreased.
  The presence of small-scale magnetic field, on the other hand, leads to smaller structure of entropy field because turbulent mixing of entropy is highly prohibited on scales smaller than the energy containing scale of the magnetic fields.
  This scale separation due to the small-scale magnetism between velocity spectrum (shift towards low wavenumber with effectively large $\nu$) and entropy spectrum (shift towards high wavenumber for effectively small $\kappa$) can be roughly modeled as an increase in the effective Prandtl number $\mathrm{Pr}$.
\begin{enumerate}
  \setcounter{enumi}{1}
  \renewcommand{\labelenumi}{(\Roman{enumi}).}
\item
  Owing to a decrease in the effective thermal diffusivity $\kappa$, low entropy is conveyed almost adiabatically to the bottom, which results in the enhancement and extension of the subadiabatic layer near the base.
\end{enumerate}
Note that the subadiabaticity here is not strong enough to be an overshoot region where downflows are quickly decelerated by buoyancy and therefore the enthalpy flux becomes downward.
The enthalpy flux is in fact transported upward even inside the subadiabatic layer formed in the lower portion of the numerical domain, which means that this layer results from the nonlocal heat transport by strong downflow plumes \citep{brandenburg2016,kapyla2017b}.
\begin{enumerate}
  \setcounter{enumi}{2}
  \renewcommand{\labelenumi}{(\Roman{enumi}).}
\item
  Downflows are subject to weak and gradual buoyant decelerations during the descent within the subadiabatic zone before reaching to the bottom beneath which a very thin and stiff overshoot layer is supposed to exist \citep{hotta2017}.
  Therefore, convective amplitude in the deeper subadiabatic layer is selectively suppressed, which may be observed as a reduction in the low wavenumber power of the horizontal velocity near the surface.
\end{enumerate}

Although the velocity suppression mechanism described above is possible, the horizontal Prandtl number $\mathrm{Pr^{H}}$-dependence of the convective rms velocities $v_{\mathrm{rms}}$ appears to be weak and still insufficient to explain the huge discrepancies between observations and simulations.
Further work is required to investigate whether this effect saturates or not at a finite effective Prandtl number, which is quite important when trying to apply this effect to the solar convective conundrum.
Future work will also focus on the effects of rotation on the degree of velocity suppression as discussed in Section \ref{subsec:5-2} or on the applicability to the solar differential rotation problem especially to clarify whether the reduction in $v_{\mathrm{rms}}$ via the thermal effect can modify the Rossby number $\mathrm{Ro}$ greatly enough to shift the differential rotation regime from anti-solar to solar-like \citep{kapyla2014,karak2015}. \\

The authors thank an anonymous referee for helpful comments on the manuscript.
Y.B. further wish to thank Mark Miesch, Ben Brown, Mark Rast, and Sacha Brun for their thoughtful comments.
This work was supported by JSPS KAKENHI Grant Number 15H03640.
Y.B. also acknowledge financial support from the Leading Graduate Course for Frontiers of Mathematical Sciences and Physics (FMSP) of the University of Tokyo.
H.H. is also supported by MEXT/JSPS KAKENHI Grant Number JP16K17655 and JP16H01169.
Numerical computations were mostly carried out on Cray XC30 at Center for Computational Astrophysics, National Astronomical Observatory of Japan (CfCA), and in part on JAXA Supercomputer System generation 2 (JSS2).
\\

\section*{APPENDIX \\ COMPARISON WITH MHD SIMULATIONS} \label{sec:appendix}

In this paper, we focus on the thermal effects of the small-scale magnetism that it can prohibit the effective thermal diffusivity.
In fact, the high-resolution simulations of small-scale dynamo should include both the dynamical effect (direct suppression by Lorentz force) and thermal effects.
However, only few attention has been paid so far on the analysis of latter mechanism.
We therefore present as an appendix the evidence that the thermal effect indeed exists in the small-scale dynamo simulations conducted by \citet{hotta2015}.
Their numerical domain consists of a simplified horizontally$(x,y)$-periodic box extending $[0 < x,y < R_\odot]$, but a realistic solar stratification, equation of state, and radiative diffusivity or energy flux values are used in a vertical direction extending $[0.715 \ R_\odot< z < 0.96 \ R_\odot]$.

\begin{figure}[]
\figurenum{13}
\plotone{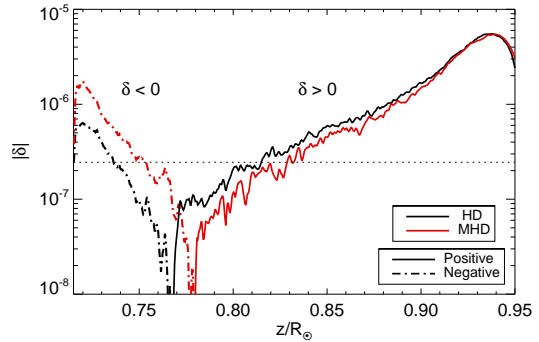}
\caption{The values of superadiabaticity obtained by small-scale dynamo simulations of \citet{hotta2015} on a vertically-logarithmic scale. The data have been averaged over $20$ days of the late simulation time. Red and black lines denote the values for calculations with and without magnetic field (corresponding to their runs H2048 and M2048). Superadiabatic and subadiabatic gradients are distinguished by solid and dot-dashed lines, respectively. Horizontal dotted line shows the value of local modified Mach number square as a typical superadiabaticity estimated by the injected energy flux in their simulations.} \label{fig:figappB}
\end{figure}

Figure \ref{fig:figappB} shows the height dependence of the superadiabaticity for their runs H2048 and M2048 calculated by,
\begin{eqnarray}
  \delta=-\frac{H_{\mathrm{p}}}{c_{\mathrm{p}}}\frac{\partial \langle s \rangle}{\partial z},
\end{eqnarray}
where $s$, $H_{\mathrm{p}}$, and $c_{\mathrm{p}}$ are the dimensional entropy, background pressure scale height, and a specific heat at constant pressure, respectively.
For making a comparison with our results easier, a typical superadiabaticity value which we use for normalizing $\delta$ is estimated as follows and overplotted in Figure \ref{fig:figappB} by a dotted line.
The typical superadiabaticity should be on the order of Mach number square, according to the mixing-length argument.
The modified Mach number at the base $M_{\mathrm{b}}$ is derived in the same way as $M_{\mathrm{b}}=v_{\mathrm{b}}/\sqrt{p_{\mathrm{b}}/\rho_{\mathrm{b}}}$, where pressure and density at the base ($z/R_\odot=0.715$) are given by $p_{\mathrm{b}}=5.79 \times 10^{13} \ \mathrm{dyn \ cm}^{-2}$ and $\rho_{\mathrm{b}}=0.195 \ \mathrm{g \ cm}^{-3}$, respectively. 
The typical convective velocity is calculated using the solar energy flux as $v_{\mathrm{b}}=(F_{\mathrm{b}}/\rho_{\mathrm{b}})^{1/3}$ with $F_{\mathrm{b}} = 1.21 \times 10^{11} \ \mathrm{erg \ s}^{-1} \mathrm{cm}^{-2}$, which finally leads to a value $M_{\mathrm{b}}^{2}=2.45\times 10^{-7}$.
Although care must be taken in that the sound speed is artificially reduced in their calculations by a factor of $150$, the comparison of the normalized superadiabaticity $\delta/M_{*,\mathrm{b}}^{2}$ would be helpful to examine the influence of the change in $\delta$.

First of all, it is obvious from Figure \ref{fig:figappB} that the subadiabatic layer formed near the base is enhanced and vertically extended with the inclusion of magnetic field.
The degree of the subadiabaticity enhancement is similar to our results with $\mathrm{Pr}=6$ Case: In both cases, the subadiabaticity near the base is increased by a factor of $2-3$.
Moreover, the vertically upward extension of the subadiabatic zone shifts the overall profile, which also leads to the decrease in the superadiabaticity in the bulk of the convection zone similarly to our results shown in Figure \ref{fig:entdel_an45}.
The rms velocity for their MHD case is then suppressed by about $60\%$ with respect to HD case (see Figure 13 of \citet{hotta2015}), which is the same suppression degree seen in our model as shown in Figure \ref{fig:vrms_an6}.
We therefore consider that in the real magnetized thermal convection, the thermal effect (velocity suppression via the enhanced subadiabatic layer due to the prohibition of the effective thermal diffusivity) may play a critical role in determining the convective amplitudes in addition to the dynamical effect.

It is also noteworthy that the superadiabaticity is not affected by the inclusion of magnetic fields near the highly-superadiabatic surface, just like our result of Case H6V1.
This, according to our discussion made in Section \ref{subsec:4-2}, indicates that the small-scale magnetic field does not inhibit the effective thermal conduction in a vertical direction: it only suppresses the ``horizontal'' effective thermal diffusivity, as already inferred by \citet{hotta2015}. 
We thus make an argument that if we try to conduct Large-Eddy Simulations with a SGS model on the effects of small-scale magnetic field, the anisotropy in the thermal diffusion term should be introduced and only the ``horizontal'' thermal diffusivity should be decreased.

\end{document}